\documentclass[aps,preprint,onecolumn,showpacs]{revtex4}

\usepackage[section]{placeins}
\usepackage{graphicx}
\usepackage{dcolumn}
\usepackage{amsmath}
\usepackage{amssymb}
\usepackage{wasysym}
\usepackage{color}
\usepackage[normalem]{ulem}
\begin{document}

\title{Nodal bilayer-splitting controlled by spin-orbit interactions in underdoped high-$T_{\rm c}$ cuprates}

\author{N.~Harrison\footnote{Correspondence to nharrison@lanl.gov}, B.~J.~Ramshaw and A.~Shekhter
}

\affiliation{Mail~Stop~E536,~Los~Alamos~National Labs.,Los~Alamos,~NM~ 87545\\
}
\date{\today}

\begin{abstract}
The highest superconducting transition temperatures in the cuprates are achieved in bilayer and trilayer  systems, highlighting the importance of intralayer interactions for high $T_{\rm c}$. It has been argued that interlayer hybridization vanishes along the nodal directions by way of a specific pattern of orbital overlap. Recent quantum oscillation measurements in bilayer cuprates have provided evidence for a residual bilayer-splitting at the nodes that is sufficiently small to enable magnetic breakdown tunneling at the nodes. Here we show that several key features of the experimental data can be understood in terms weak spin-orbit interactions naturally present in bilayer systems, whose primary effect is to cause the magnetic breakdown to be accompanied by a spin flip. These features can now be understood include the equidistant set of three quantum oscillation frequencies, the asymmetry of the quantum oscillation amplitudes in $c$-axis transport compared to $ab$-plane transport, and the anomalous magnetic field angle dependence of the amplitude of side frequencies suggestive of small effective {\it g}-factors. We suggest that spin-orbit interactions in bilayer systems can further affect the structure of the nodal quasiparticle spectrum in the superconducting phase. 
\end{abstract}
\maketitle

\section*{Introduction}
\begin{figure}[ht!!!!!!!!!!] 
\centering
\includegraphics*[width=0.8\columnwidth]{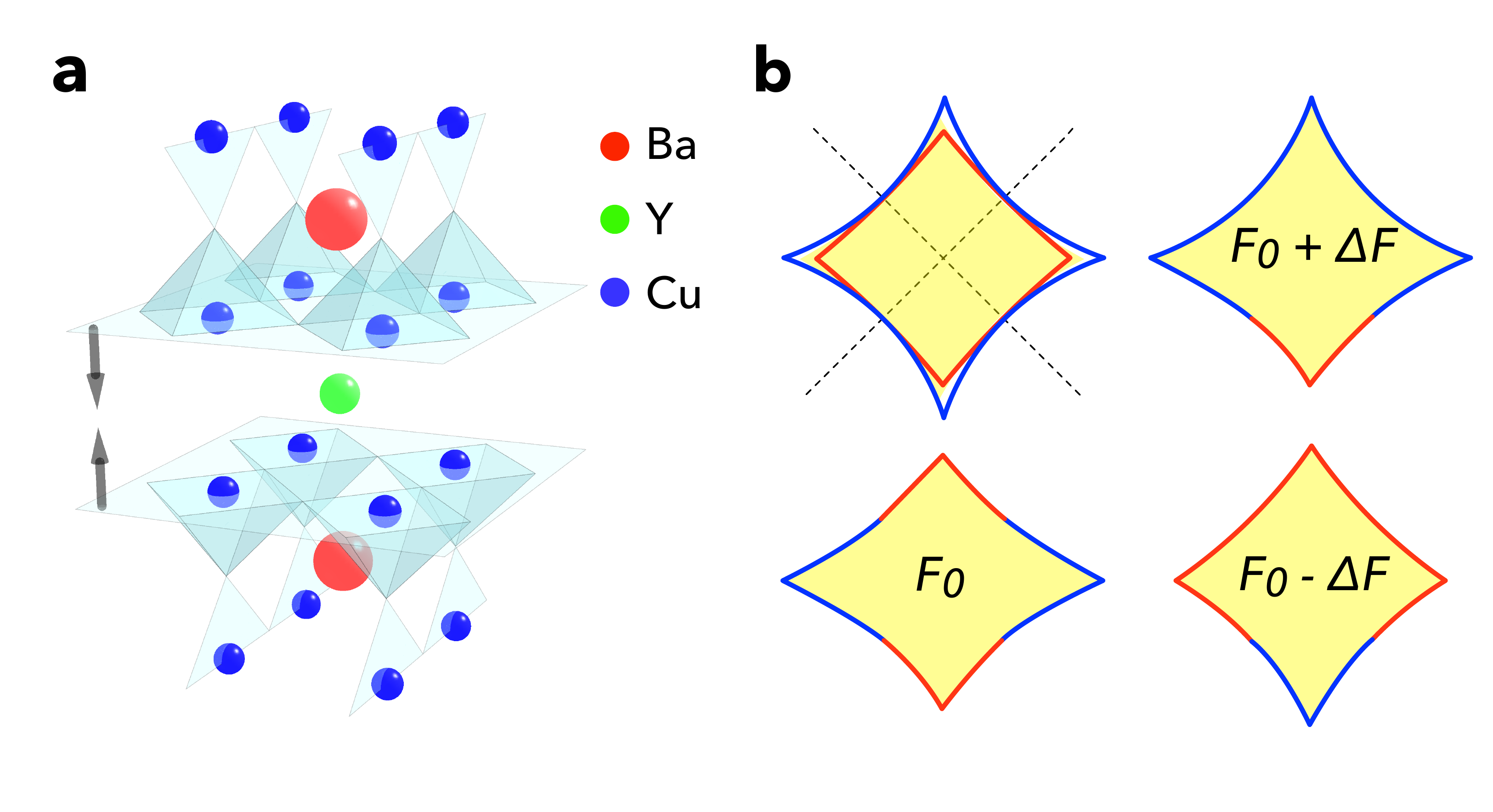}
\caption{{\bf Schematic bilayer structure of YBa$_2$Cu$_3$O$_{6+x}$.} 
({\bf a}), Schematic crystal structure of bilayer YBa$_2$Cu$_3$O$_{6+x}$. The non-centro-symmetric environment of each CuO$_2$ plane is captured with a polar vector (indicated by arrows) which has  opposite directions in each layer due to overall inversion symmetry. ({\bf b}), Schematic bilayer splitting of the proposed electron pocket~\cite{sebastian1,sebastian2} into bonding (blue) and antibonding (red) pockets, together with a schematic of some of the magnetic breakdown combination orbits resulting from a small residual bilayer-splitting along the nodal directions (indicated by dotted lines).}
\label{bilayer}
\end{figure}

The hybridization between multiple copper-oxide layers is an important factor in achieving high superconducting transition temperatures in the copper-oxide materials~\cite{legget1} (see Fig.~\ref{bilayer}a for bilayer crystal structure of YBa$_2$Cu$_3$O$_{6+x}$). The bilayer-splitting of the energy spectrum into bonding and antibonding bands caused by this hybridization is directly observed in photoemission experiments on Bi$_2$Sr$_2$CaCu$_2$O$_{8+x}$~\cite{kordyuk1} and YBa$_2$Cu$_3$O$_{6+x}$~\cite{fournier1}. The quasiparticle energy dispersion along the $<$110$>$ (or nodal) direction is of particular interest owing to the predicted vanishing of the bilayer splitting along this direction~\cite{andersen1}. In YBa$_2$Cu$_3$O$_{6+x}$, the nodal bilayer splitting has been shown to fall below the resolution limit along the nodal direction for hole dopings below $\approx$~15\%~\cite{fournier1}---the doping below which quantum oscillations are observed. In Bi$_2$Sr$_2$CaCu$_2$O$_{8+x}$, it remains small at all dopings~\cite{kordyuk1}. In contrast to graphene, where the degeneracy between two bands at high symmetry points in the Brillouin zone is protected by the crystalline symmetry,  the degeneracy points of the quasiparticle spectra in bilayer copper oxides are located at a point of lower symmetry in unreconstructed Brillouin zone. This suggests that the quasiparticle spectrum in the nodal region of bilayer copper oxides is a sensitive indicator of small effects, such as higher order hopping terms, spin-orbit interactions and electronic correlations~\cite{maharaj1,hosur1,garcia1,fournier1,kordyuk1,andersen1}, even when no further symmetry breaking terms are present.

The discovery of magnetic quantum oscillations in the underdoped copper oxide superconductors~\cite{doiron1,barisic1,yelland1,bangura1} has opened up a new route towards understanding the metallic state in underdoped high temperature superconducting cuprates. After several years of research, the emerging Fermi surface picture is one of a single electron pocket per CuO$_2$  plane~\cite{harrison1,sebastian9,doiron2,maharaj1,allais1,zhang1,tabis1,senthil1}, that is composed of the nodal regions of the original large unreconstructed Fermi surface~\cite{andersen1}. Such a picture brings together a range of experimental observations, including quantum oscillations, x-ray scattering evidence for broken translational symmetry~\cite{ghiringhelli1,chang1}, the negative Hall coefficient~\cite{leboeuf1} and the high magnetic field electronic heat capacity~\cite{riggs1}. Bilayer splitting of the small pocket is expected to manifest itself as multiple quantum oscillation frequencies, and multiple frequencies have indeed been revealed by the presence of a beat pattern in quantum oscillation measurements made on the bilayer cuprate YBa$_2$Cu$_3$O$_{6+x}$ by multiple groups ~\cite{audouard1,ramshaw1,sebastian1,sebastian2}. The magnetic field-dependent quantum oscillation amplitude is not zero at the nodes of the beating pattern (see Fig.~\ref{data}a), suggesting the presence of at least three frequencies. Experiments clearly identify three oscillation frequencies -- a dominant frequency at $F_0\approx$~530~T and two side frequencies at $F_- = F_0-\Delta F\approx$~440~T and $F_+=F_0+\Delta F\approx$~620~T~\cite{sebastian1,sebastian2,ramshaw1,audouard1}. 

The equidistant spacing of the three observed frequencies, $F_-, F_0$  and $F_+$, suggests a common origin for their corresponding orbits (as opposed to the orbits existing in different parts of the Brillouin zone). Combination orbits resulting from magnetic breakdown in which the tunneling of electrons across a small residual gap separating bilayer-split Fermi surfaces (see Fig.~\ref{bilayer}b) can naturally account for such an equidistant set of frequencies~\cite{sebastian1,sebastian2}. The spacing $\Delta F$ is directly related to the difference in area between bonding- and antibonding-hybridized Fermi surface pockets in such a case. A common origin is further supported by similar angular dependences of the two side frequencies $F_-$ and $F_+$. A close examination of the angle-dependent measurements of quantum oscillations in YBa$_2$Cu$_3$O$_{6+x}$ (see Fig.~\ref{data}) suggests an anomalously small effective {\it g}-factor for these side frequencies~\cite{sebastian2}, which is atypical for a 3{\it d} transition metal system.

The vanishing bilayer hybridization along the nodal directions~\cite{andersen1} leaves open a number of possible mechanisms for magnetic breakdown tunneling between bilayer-split Fermi surfaces along the nodal direction. Here we show that spin-orbit interactions, naturally present in bilayer crystalline systems, provide just such a mechanism. It enables an understanding of the observed anomalously small effective {\it g}-factors of side frequencies, as well as several other key features of the experimental quantum oscillation data.

\section*{Results}
\subsection*{Angle-dependent quantum oscillation amplitude in YBa$_2$Cu$_3$O$_{6+x}$ originating from Zeeman splitting}
Interference between two linearly Zeeman-split components of a cyclotron orbit in a magnetic field suppresses the quantum oscillation amplitude by a factor
\begin{equation}\label{spindamping}
R_{\rm s}=\cos\bigg[\frac{\pi m{g}_{\rm eff}}{2m_{\rm e}\cos\theta}\bigg]
\end{equation}
that depends on the ratio of Zeeman energy $g_{\rm eff}\mu_{\rm B}B$ to the cyclotron energy $\hbar\omega_{\rm c}=\hbar eB\cos\theta/m$~\cite{shoenberg1}. Here $B=|{\bf B}|$, $g_{\rm eff}$ is an effective {\it g}-factor, $m$ is the cyclotron effective mass in a quasi-two-dimensional metal, $m_{\rm e}$ is the free electron mass while $\theta$ is the angle between ${\bf B}$ and the crystalline $c$-axis. Experiments have shown that $m \approx$~1.6~$m_{\rm e}$ in YBa$_2$Cu$_3$O$_{6+x}$ at $x\sim$~0.6~\cite{ramshaw1,sebastian9}, while $g_{\rm eff}$ is generally expected to be renormalized by factors different from those renormalizing $m$~\cite{shoenberg1}.
Since both copper and oxygen have only very weak intrinsic spin-orbit interactions in YBa$_2$Cu$_3$O$_{6+x}$~\cite{walstedt1}, the Zeeman splitting of the Fermi surface is weakly dependent on the direction of the magnetic field. The cyclotron frequency, by contrast, scales with $1/\cos\theta$ owing to the cylindrical geometry of the Fermi surface. As we rotate the field into the copper-oxide plane, $R_{\rm s}$, given by Equation~\ref{spindamping}, goes through a series of `spin zeroes,' which occur when $(m/m_{\rm e})g_{\rm eff}/\cos\theta$ is equal to an odd integer~\cite{shoenberg1}. 

\begin{figure}[ht!!!!!!!!!!]
\centering 
\includegraphics*[width=0.55\columnwidth]{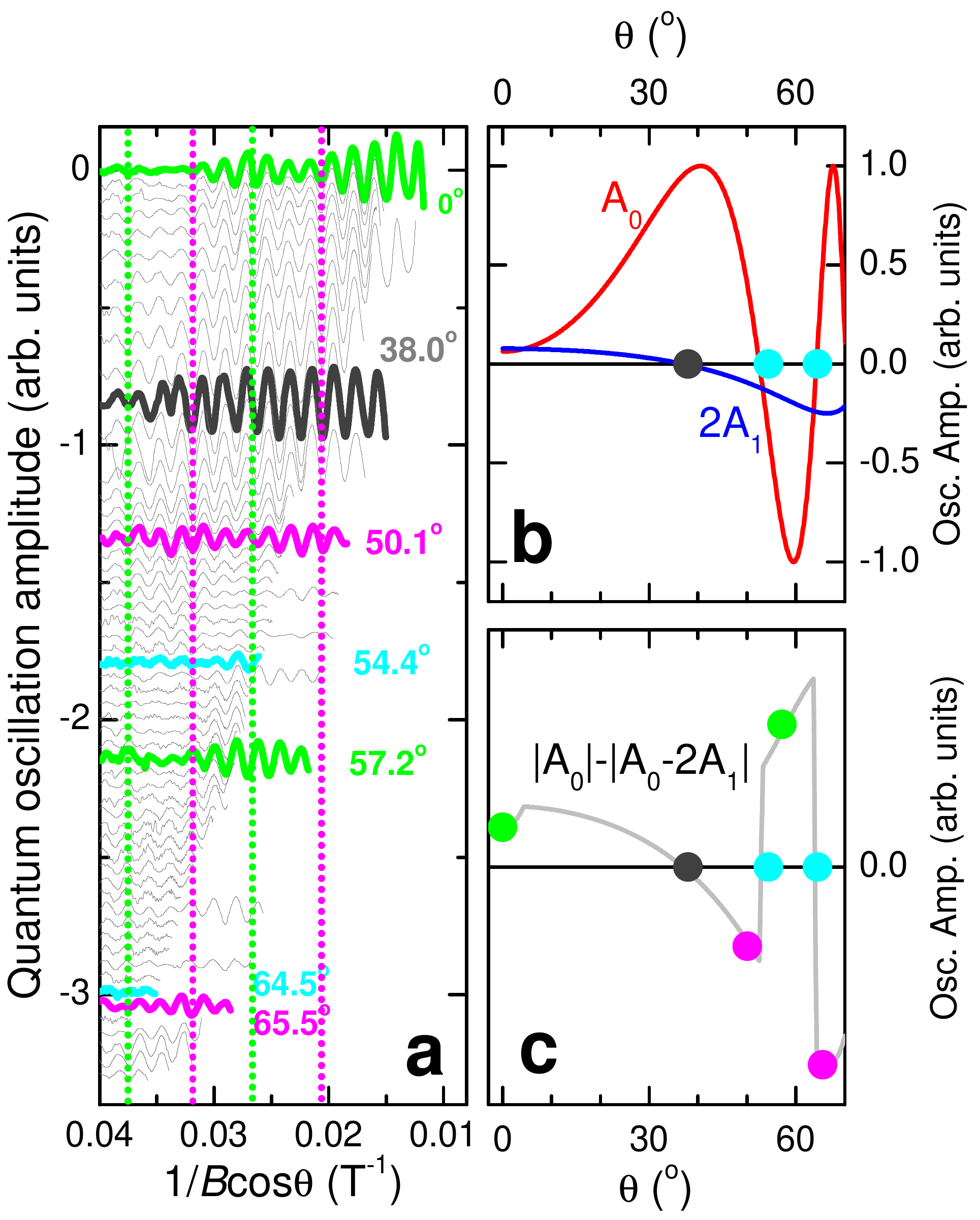}
\caption{{\bf Angle-dependent quantum oscillations in YBa$_2$Cu$_3$O$_{6+x}$.}
({\bf a}), Quantum oscillations from Ref.~\cite{sebastian2} (thin grey lines). Prominent beat patterns of the same phase 
at two very different angles $\theta$ are highlighted in green. Sweeps for which the beat pattern is maximally phase shifted relative to these are highlighted in magenta. Curves lying closest to the spin zeroes in $A_0$ are highlighted in cyan, while that lying closest to the spin zero in $A_1$ (i.e. $A_-$ and $A_+$) is highlighted in dark grey. Vertical lines are drawn to assist viewing the relative phase of the beat. ({\bf b}) Simulated $R_{\rm s}$ for $A_0$ and $2A_1$ (the latter assumed to be of smaller amplitude as in the experiments), approximately representing the $\theta$-dependent amplitudes of the central $F_0$ (red) and side $F_-$ and $F_+$ (blue) frequencies, respectively. We use $g_{\rm eff}=$~1.9 for $A_0$ and $g_{\rm eff}=$~0.5 for $A_1$, and $m=$~1.6~$m_{\rm e}$ for both. Colored circles indicate the angles of the highlighted sweeps in ({\bf a}). ({\bf c}) Amplitude (grey line) $|A_0-2A_\pm|-|A_0|$ of the oscillatory component of the quantum oscillation envelope according to $A_0$ and $2A_1$ in ({\bf b}). For clarity, the vertical locations of the green and magenta circles are chosen so as to lie on the curve.}
\label{data}
\end{figure}

The amplitude $A_{0}$ of the dominant quantum oscillation frequency ($F_0$) in YBa$_2$Cu$_3$O$_{6+x}$ has been shown to cross zero at $\approx$~53.9$^\circ$~\cite{ramshaw1,sebastian2} and $\approx$~64.7$^\circ$--- the higher of these angles being accessible only under very high magnetic fields owing to the increase in the superconducting upper critical field with angle. The closest experimental magnetic field sweeps to both these angles are indicated in cyan in Figs.~\ref{data}a, b and c. These two spin zeroes can be consistently understood in terms of an effective {\it g}-factor of $g_{\rm eff}\approx$~2 for the main frequency $F_0$~\cite{ramshaw1,sebastian2}, which is similar to that found in many non correlated metals~\cite{shoenberg1}.

\subsection*{Anomalous Zeeman splitting of the side frequencies $F_-$ and $F_+$}
The side frequencies are responsible for a beat pattern in the overall oscillation amplitude~\cite{audouard1,sebastian1}. The total quantum oscillation waveform can be represented as a sum $-A_0\cos\frac{2\pi F_0}{B}-A_-\cos\frac{2\pi{F}_-}{B}-A_+\cos\frac{2\pi{F}_+}{B}$, or, alternatively, as a product $\big[A_0+2A_1\cos\frac{2\pi\Delta F}{B}\big]\cos\frac{2\pi F_0}{B}-2A_2\sin\frac{2\pi\Delta F}{B}\sin\frac{2\pi F_0}{B}$. This describes oscillations of the main frequency $F_0$ modulated by an envelope function, where $A_{1,2}=\frac{1}{2}(A_-\pm{A}_+)$. To leading order in $A_1/A_0$ , the envelope function is given by $|A_0+2A_1\cos(\frac{2\pi\Delta F}{B})|-|A_0|$. The amplitude of the slowly oscillating component of the envelope is therefore $|A_0+2A_1|-|A_0|$.

As we rotate the magnetic field away from the $c$-axis, spin zeroes in both $A_{0}$ and $A_1$ lead to zero-crossings of the beat amplitude $|A_0+2A_1|-|A_0|$. The data shows only three angles where this occurs, of which two are associated with spin zeroes in $A_0$ while only one at $\approx$~38$^\circ$ (highlighted in dark grey in Fig.~\ref{data}a) is associated with a spin zero in $A_1$ (see Fig.~\ref{data}a). The spin zero in $A_1$ where the oscillatory component of the beat vanishes is further accompanied by a very large amplitude in $A_0$~\cite{ramshaw1}. Only one spin-zero in $A_1$ over such a broad angular range requires an anomalously small effective {\it g}-factor associated with the $F_-$ and $F_+$ orbits~\cite{sebastian2}.

On assuming simple forms for $R_{\rm s}$ in Fig.~\ref{data}b, the angle-dependence of the beat amplitude $|A_0+2A_1|-|A_0|$ obtained in Fig.~\ref{data}c can be seen to qualitatively account for the angle-dependence of the beat pattern seen in the experimental data in Fig.~\ref{data}a. The experimental data are therefore consistent with values of $g_{\rm eff}\approx$~2  for the central frequency ($F_0$) and $g_{\rm eff}\approx$~0.5 for the side frequencies ($F_-$ and $F_+$). We note that the $R_{\rm s}$ factors are the only source of strong angular dependences that can cause sign changes in $A_0$ and $A_1$ on varying $\theta$. 
Factors such as the amplitude prefactor and the thermal, Dingle and magnetic breakdown amplitude reduction factors~\cite{shoenberg1} are independent of $\theta$ for fixed $B\cos\theta$. Similarly, chemical potential oscillations have been shown to not significantly affect the amplitude of the fundamental over the entire angular range~\cite{sebastian9}. Although Fermi surface warping does lead to an additional angular dependence, its effect is relatively weak owing to the small area of the pocket~\cite{bergemann1,sebastian2,ramshaw1}. 

It is important to note that because $R_{\rm s}$ depends on the product $mg_{\rm eff}$~\cite{shoenberg1} (not just on $g_{\rm eff}$), differences in effective mass between orbits need careful consideration. While very different effective masses could occur for orbits originating from different regions of the Brillouin zone, the combination orbits are expected to have similar effective masses owing to the small magnitude of the bilayer splitting compared to the Fermi energy. Detailed studies of the temperature-dependent wave form of the quantum oscillations have suggested an upper bound for the difference in effective mass of $\Delta m\lesssim$~0.2~$m_{\rm e}$~\cite{sebastian9,ramshaw1} for different frequencies, which is a small fraction of $m$. The four times smaller value of $mg_{\rm eff}$ in the argument of $R_{\rm s}$ for the $F_-$ and $F_+$ frequencies compared to that of $F_0$ must therefore be attributed to a small value of $g_{\rm eff}$ for these frequencies. This further suggests that magnetic breakdown must be responsible for producing anomalously small effective {\it g}-factors for the side frequencies.  We note that while renormalizations due to interactions can lead to different effective masses entering the $R_{\rm s}$ factors and the thermal damping factors, the value of $g_{\rm eff}\approx$~2 obtained for the main frequency suggests that this difference is beyond the present experimental resolution.

\subsection*{Spin-orbit interactions in bilayer cuprates and anomalous {\it g}-factors.}
The effective {\it g}-factor quantifies the absolute rate-of-change in the cyclotron orbit area with applied magnetic field. In the limit of small spin-orbit interactions, the orbit itself can consist of expanding segments for which the spin is aligned parallel to the increasing applied magnetic field and shrinking segments for which the spin is anti-aligned. A small effective {\it g}-factor $g_{\rm eff}\sim$~0 corresponds to nearly equal time spent in each polarization state on completing a cyclotron orbit, while a large $g_{\rm eff}=2$ corresponds to a fixed polarization state throughout the orbit. The observed anomalously small effective {\it g}-factor for the side frequencies in YBa$_2$Cu$_3$O$_{6+x}$ implies that the spin polarization state must change as the electrons travel along their corresponding cyclotron orbits. This can occur in a network of coupled orbits~\cite{shoenberg1} if an effective mechanism exists for spin-flips to occur while traversing the orbit, and if semiclassical trajectories on the Fermi surface corresponding to opposite spin projections intersect each other in the momentum-space. While a broken time-reversal symmetry phase, such as a spin-density wave~\cite{norman1}, could be considered as a mechanism for producing effective spin-flips, such a possibility has been ruled out experimentally in YBa$_2$Cu$_3$O$_{6+x}$~\cite{wu1}.  In the absence of spin-density waves, weak spin-orbit interactions associated with the lack of inversion symmetry in each CuO$_2$ plane of the copper oxide bilayer~\cite{aji1} provide a viable mechanism by which spin-flips can occur (see schematic in Fig.~\ref{bilayer}a). The observed value of $g_{\rm eff}\approx$~2 for the dominant frequency implies that spin-orbit interactions are weak. The spin and momentum are therefore not locked except near the nodal regions where bilayer splitting causes the semiclassical trajectories of opposite spin polarization to intersect. It is only here that spin-orbit interactions can have a significant effect on the electron kinematics. 

Spin-orbit interactions within a bilayer of CuO$_2$ can be treated using the standard bilayer Hamiltonian 
\begin{equation}\label{bilayerhamiltonian}
H_{\rm bilayer}=\left( \begin{array}{cc}H_{\rm layer}^+&t_{\perp,{\bf k}}\sigma_0\\t_{\perp,{\bf k}}\sigma_0&H_{\rm layer}^-\end{array} \right)
\end{equation}
where the intralayer Hamiltonian $H_{\rm layer}^\pm=\varepsilon_{\bf k}\sigma_0\pm\alpha(\boldsymbol{\sigma}\times{\bf k})\cdot\boldsymbol{\hat{z}}+\frac{1}{2}g\mu_{\rm B}({\bf B}\cdot\boldsymbol{\sigma})$ represents kinematics inside each each layer and $t_{\perp,{\bf k}}\approx t_{\perp}(\cos a {\bf k}_x - \cos b {\bf k}_y )^2$ is the intra-bilayer hybridization~\cite{andersen1}. Here  $g$ is the electron {\it g}-factor, $\boldsymbol{\sigma}=(\sigma_x,\sigma_y,\sigma_z)$ are  Pauli matrices and $\sigma_0$ is the 2~$\times$~2 identity matrix. The lack of an inversion center in the chemical environment for each layer can be represented by an out-of-plane polar vector ($\boldsymbol{\hat{z}}$ in Figure 1), which gives rise to Rashba-type spin-orbit interactions, $\alpha(\boldsymbol{\sigma}\times\boldsymbol{k})\cdot\hat{\boldsymbol{z}}$ ~\cite{bychkov1}. The overall inversion symmetry of the bilayer requires that $\boldsymbol{\hat{z}}$ has opposite directions in the two layers (see Figure 1a), which we represent here by opposite values of  $\alpha$. A small orthorhombic distortion  of YBa$_2$Cu$_3$O$_{6+x}$ along the $a$ or $b$ axes can also allow the `nematic' terms $\beta\eta(\sigma_xk_x-\sigma_yk_y)$ where $\boldsymbol{\eta}$  transforms under $B_{1u}$ . Here again, due to overall inversion symmetry of the bilayer,  $\beta$ has opposite values in the two layers. We note that these two forms of spin-orbit interaction do not exhaust all possibilities in the tetragonal group of YBa$_2$Cu$_3$O$_{6+x}$. 

\begin{figure}[ht!!!!!]
\centering
\includegraphics[width=1\columnwidth]{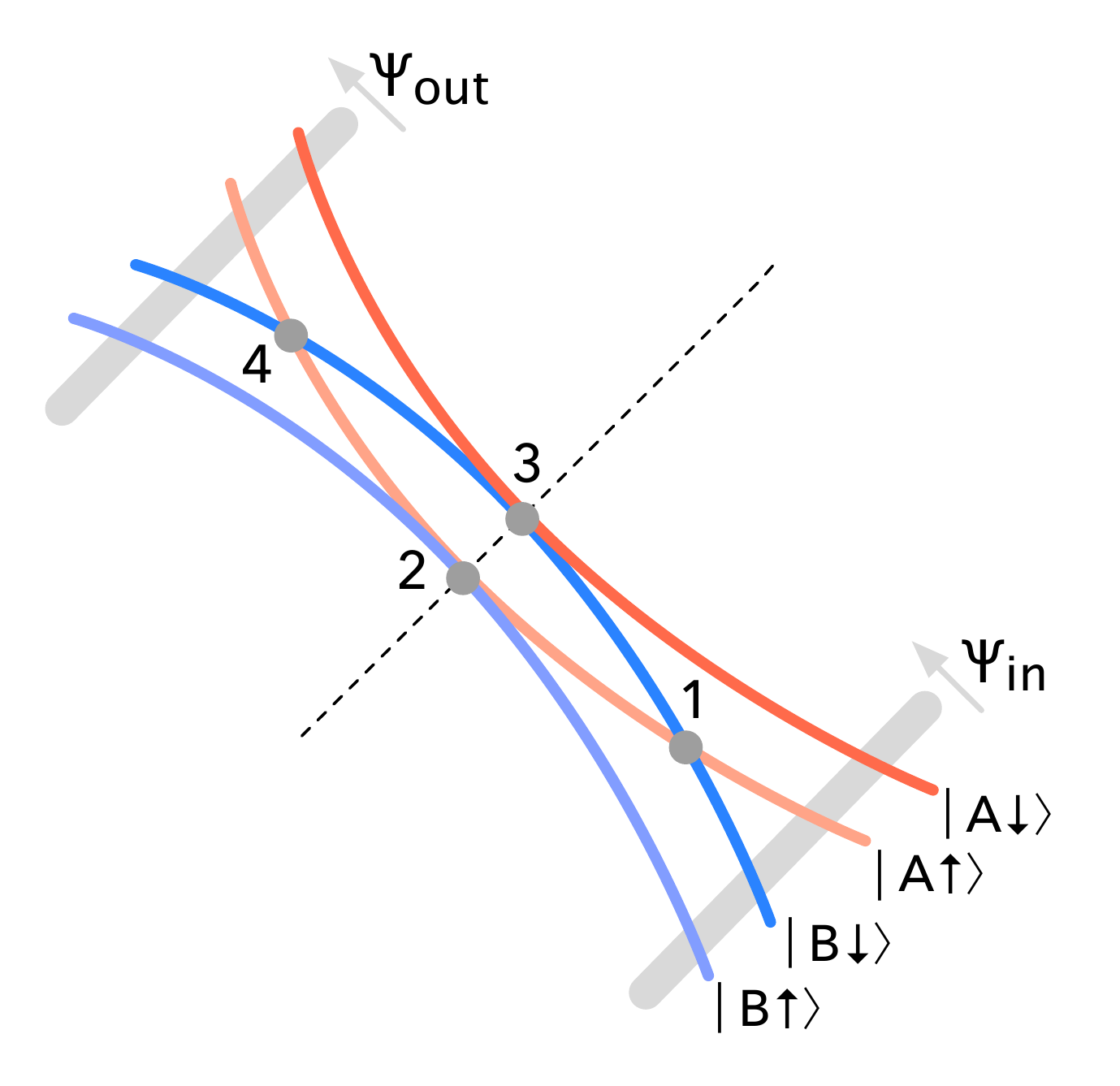}
\caption{{\bf Schematic of the semiclassical trajectories in the nodal Fermi surface region}. 
The trajectories are shown for vanishing spin-orbit interactions at $B\neq0$, with the node in $t_{\perp,{\bf k}}$ indicated by a dashes line. The semiclassical magnetic breakdown tunneling junctions are indicated numerically.
}
\label{trajectories}
\end{figure}

Weak spin-orbit interactions can only cause spin-flips at the nodal region of the orbit where $t_{\perp,{\bf k}}$ vanishes~\cite{andersen1,garcia1} (see Fig.~\ref{trajectories} for a schematic of the nodal region), allowing semiclassical trajectories with opposite spin projections to cross paths. Note that the semiclassical trajectories will not cross in momentum-space if $t_{\perp,{\bf k}}$ is too large at the nodes. While a spin-flip can still occur in this situation, its probability will be exponentially suppressed. Far from the nodal regions, electrons propagate along trajectories with a well defined effective bilayer parity `A' and `B' (i.e. behavior under reflection in the mid-plane of the bilayer, where `B' stands for bonding and `A' stands for antibonding) and spin projection ($\uparrow$ for spin-up and $\downarrow$ for spin-down). The basis for propagation between nodes can be represented by a four-component spinor 
\begin{align}
\Psi=& (\psi_{{\rm B}\uparrow}, \psi_{{\rm B}\downarrow}, \psi_{{\rm A}\uparrow}, \psi_{{\rm A}\downarrow})\,. \notag
\end{align}
As the nodal region is traversed, weak spin-orbit interactions and sub-leading hopping terms introduce mixing between states (see methods), which we describe using a 4$\times$4 transfer matrix $\hat{T}$, where 
\begin{align}\label{transfer}
\Psi_{\rm out}=\hat{T}\Psi_{\rm in}
\end{align}
and $\Psi_{\text{in, out}}$ refer to the incoming and outgoing electron states, respectively (see Fig.~\ref{trajectories}).

In the limit where the Zeeman energy is small compared to the bilayer coupling at its maximum in the antinodal directions, the combination of four different states for each orbit quadrant give rise to 3 possible integer values for $g_{\rm eff}$ on completing an orbit (see Fig.~\ref{orbitsubset} for examples). Orbits comprising 2 spin-up and two spin-down quadrants have $g_{\rm eff}\approx$~0, those comprising 3 spin-up and 1 spin-fown quadrants or {\it vice versa} have $g_{\rm eff}\approx$~1 while those comprising 4 spin-up or 4 spin-down quadrants have $g_{\rm eff}=$~2.

\subsection*{Angle-dependence of the quantum oscillation amplitude}
The experimental quantum oscillation amplitudes $A_0$, $A_-$ and $A_+$ depend critically on the structure of the transfer matrix $\hat{T}$ in the vicinity of the nodal directions. While a more realistic description of the current experimental situation must take into consideration the effects of Landau quantization and the near-degeneracy of the four trajectories in the nodal region, the essential features of the experimental data are captured by a simple semiclassical anzatz for $\hat{T}$. For simplicity, this anzatz neglects the finite momentum-space extent of the wave functions (especially at the low Landau level indices relevant to experiments~\cite{sebastian9}) and vanishingly small separation between orbits over an extended region of momentum space~\cite{khodas1}, which can cause the transfer through the nodal region to be non-local (i.e., distributed over a finite region in momentum space).    
In this approximation, the transfer through the nodal region occurs at a distinct set of junctions (1, 2, 3 and 4 in Figures~\ref{trajectories} and \ref{zoom}b)~\cite{shoenberg1}, which enables a factorization of the transfer matrix according to $\hat{T}= \hat{T}^{\dagger}_{\rm SO}\hat{F}_\Phi\hat{T}_{\text{BA}}\hat{F}_\Phi\hat{T}_{\rm SO}$, where $\hat{T}_{\rm BA}$ describes transfers through junctions 2 and 3, and the matrices $\hat{T}_{\text{SO}}$ and $\hat{T}^{\dagger}_{\text{SO}}$ describe transfers through junctions 1 and 4 (complete matrices given in the methods). $\hat{F}_\Phi$ is a diagonal matrix desribing the free cyclotron motion between junctions. It is convenient to adopt the amplitude and phase notation used in connection with magnetic breakdown combination orbits~\cite{shoenberg1} (i.e. $p$ and $q$). Spin-flips occur only at junctions 1 and 4 with amplitude $q_1$. This matrix element describes a combined bilayer parity and spin switching transition from B$\downarrow$ to A$\uparrow$. The non spin-flip bilayer parity-conserving transition has an amplitude $ip_1$ constrained by $p_1^2+q_1^2 =1$. The transfer through junctions 2 and 3, meanwhile, affects only the bilayer parity index. This has the same amplitude $ip_2$ for the transition from B$\uparrow$ to A$\uparrow$ and from B$\downarrow$ to A$\downarrow$. The spin and bilayer parity conserving amplitude $q_2$ from B$\uparrow$ to B$\uparrow$, etc., again, is constrained by $p_2^2+q_2^2=1$.

\begin{figure}[ht!!!!!]
\centering
\includegraphics[width=0.9\columnwidth]{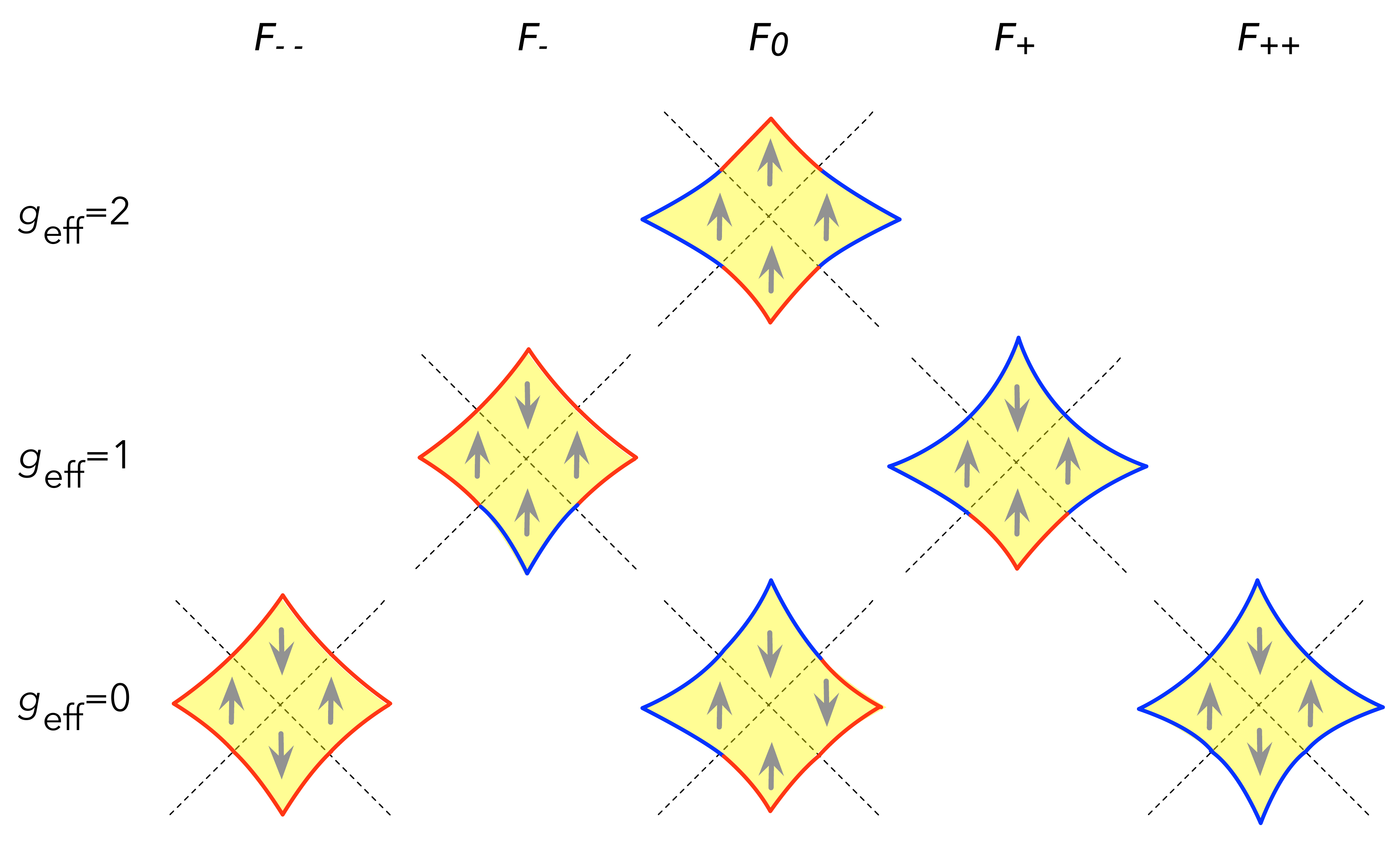}
\caption{{\bf Dominant combination orbits in the quasiclassical approximation for the transfer matrix $\hat{T}$}. In each quadrant the bilayer parity (`A' or `B') is indicated by color (red or blue) and spin is indicated by arrow. 
The $F_0$ frequency for which $g_{\rm eff}=2$ has amplitude $NR_{\rm MB}=2p_1^4$ (where $NR_{\rm MB}$ is the product of the number of times the same orbit is repeated in the Brillouin zone and the magnetic breakdown amplitude factor), while that for which $g_{\rm eff}=0$ has amplitude $NR_{\rm MB}=-4p_1^2q_1^2$.
The $F_-$ and $F_+$ frequencies have amplitude $NR_{\rm MB}=-4p_1^2q_1^2$, while the $F_{--}$ and $F_{++}$ frequencies have amplitude $NR_{\rm MB}=q_1^4$.}
\label{orbitsubset}
\end{figure}

\begin{figure}[ht!!!!!]
\centering
\includegraphics[width=0.5\columnwidth]{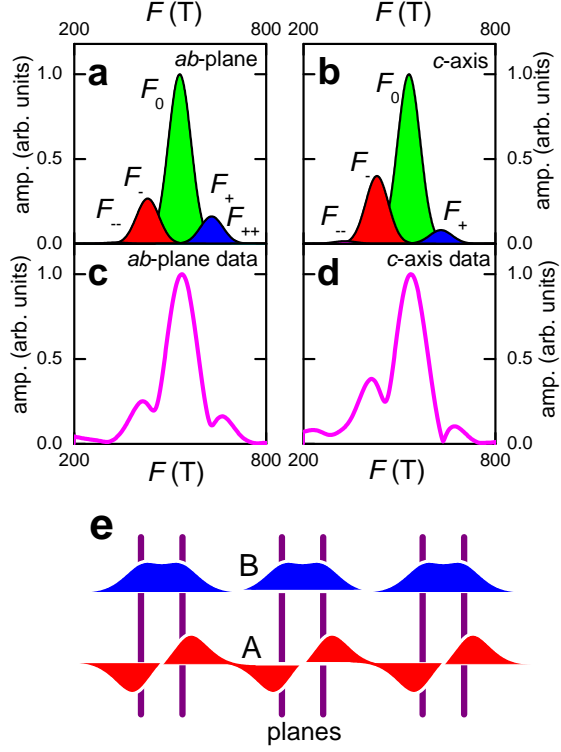}
\caption{{\bf Illustration of the effects of bonding-antibonding band asymmetry}. 
({\bf a}), Schematic Fourier amplitudes of the the $F_{--}$, $F_-$, $F_0$, $F_+$ and $F_{++}$ frequencies in in-plane transport at $B=$~44~T, for $q_1=$~0.28 and $q_2=$~0, computed using $a_{\rm in-plane}=NR_{\rm MB}R_{\rm D}$ (valid for in-plane transport), where $NR_{\rm MB}$ is the magnetic breakdown amplitude (see Fig.~\ref{orbitsubset}) and $R_{\rm D}=e^{-\frac{\Gamma}{B}\sqrt{\frac{F}{F_0}}}$ is a Dingle damping factor estimated for a constant mean free path. The latter causes the amplitude of the larger $F_+$ frequency to be slightly lower than that of the $F_-$ frequency. The overall amplitude has been renormalized so that $F_0$ has unity amplitude. ({\bf b}), Schematic of the Fourier amplitudes expected for the $c$-axis transport computed using $a_{c-{\rm axis}}=\big(\frac{t}{t_0}\big)NR_{\rm MB}R_{\rm D}$. Here, $\frac{t}{t_0}$ weights the amplitude in proportion to the fraction of orbit time spent in the antibonding band (assuming the antibonding band to dominate the $c$-axis conductivity) causing $F_-$ to have a significantly greater amplitude than $F_+$. ({\bf c}), Fourier transform in $1/B$ of contactless in-plane transport quantum oscillations measured in YBa$_2$Cu$_3$O$_{6.59}$ for a field interval 33~$\leq B\leq$~65~T at $T=\approx$~1.5~K. ({\bf d}), Equivalent Fourier transform for $c$-axis transport measured on the same sample as in ({\bf c}). 
 ({\bf e}), Schematic of the bonding (B) and antibonding (A) band orbitals (in blue and red respectively) along the $c$-axis (drawn horizontally), with the bilayers represented by purple lines. 
}
\label{FFT}
\end{figure}

Within the semiclassical approximation, experimental observations place constraints on $p_1$, $q_1$, $p_2$ and $q_2$. The observation of spin-zeroes in $A_-$ and $A_+$ (see Fig.~\ref{data}) implies that the spin-up and spin-down contributions to each frequency must be similar in amplitude, which can only occur in the limit $q^2_2/p^2_2\rightarrow0$. Such a limit corresponds to a near unity tunneling probability between bonding and antibonding bands at the nodal junctions 2 and 3, which has the effect of greatly reducing the number of experimentally relevant orbits. Meanwhile, the observation of a large amplitude for the dominant $F_0$ frequency implies that $p_1>q_1$.
Figure~\ref{orbitsubset} shows examples of the subset of orbits that have significant amplitudes in the limit $q^2_2/p^2_2\rightarrow0$. A value of $q_1\sim$~0.3 produces a series of frequencies with relative amplitudes in qualitative agreement with experiments (see Fig.~\ref{FFT}). 

\begin{figure}[ht!!!!!]
\centering
\includegraphics[width=0.7\columnwidth]{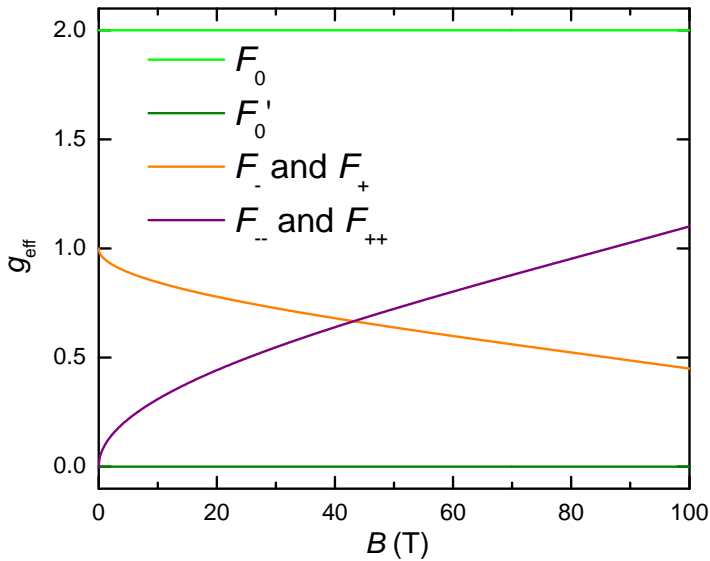}
\caption{{\bf Magnetic field-dependent effective {\it g}-factors within the semiiclassical approximation}. 
The magnetic field-dependence illustrated here arizes from the shifting of the junctions 1 and 4 away from the nodal direction. See text for other possible sources of angle and field-dependences of the effective g-factor. 
We consider $t_{\perp,{\bf k}}=t_{\perp,\phi=90^{\circ}}\cos^2(2\phi)$ and $t_{\perp,\phi=90^{\circ}}=$~10~meV.  $\psi_{{\rm A}\uparrow}$ and $\psi_{{\rm B}\downarrow}$ intersect when $t_{\perp,0}\cos^2(\frac{\pi}{2}-2\Delta\phi)=h$. In the case of the $F_-$ and $F_+$ frequencies, $g_{\rm eff}\approx1-\frac{4\Delta\phi}{\pi}$, while in the case of the $F_{--}$ and $F_{++}$ frequencies, $g_{\rm eff}\approx\frac{8\Delta\phi}{\pi}$.
}
\label{gfactors}
\end{figure}

Whereas $g_{\rm eff}=$~1 for the side frequencies $F_-$ and $F_+$ in the limit where the Zeeman interaction is small compared to the bilayer hybridization, $g_{\rm eff}$ is expected to decrease in strong magnetic fields. Under magnetic fields of the strength required to see quantum oscillations ($B\gtrsim$~20~T),
the tunneling junctions 1 and 4 move away from the nodal lines in momentum-space (shown schematically in Figs.~\ref{trajectories} and \ref{zoom}), causing the sections of the cyclotron orbit with spins aligned either parallel or opposite to the magnetic field to no longer have equal areas. The angle $\Delta\phi$ in momentum-space subtended by the arc spanning junctions 1 or 4 and the nodal line grows with magnetic field, causing the effective {\it g}-factor for some of the orbits to acquire different values at high magnetic fields (see Fig.~\ref{gfactors}). The effective {\it g}-factors of the $F_-$ and $F_+$ frequencies are found to approach a value of $\sim$~0.5 at $B\approx$~80~T, which is consistent with the observation of a single spin-zero for these frequencies at $\approx$~38$^\circ$ in experiments. The large amplitude of $A_0$ compared to $A_1$ and the slow modulation of the beat, however, precludes a reliable analysis of the magnetic field dependence of the angle at which this spin zero occurs. 
We note that additional angle and magnetic field-dependences of the effective {\it g}-factors could also arise from the magnetic field and angle-dependences of $\hat{T}_{\rm BA}$ and $\hat{T}_{\rm SO}$, which, in the semiclassical approximation, will be captured by the magnetic field and angle-dependences of $p_1$, $q_1$, $p_2$  and $q_2$. 

\subsection*{Hybridization effects in the bonding and antibonding bands}
Thus far we have neglected the orbital character of the bonding and antibonding bands, which may cause a difference in effective mass, a difference in quasiparticle scattering rate or a difference in the strength of $c$-axis hopping. A difference $\Delta m$ in effective mass~\cite{sebastian9,ramshaw1} is expected to produce effective masses for the $F_{--}$, $F_-$, $F_0$, $F_+$ and $F_{++}$ frequencies in the form of an arithmetric series in which the increment is $\Delta m$. On considering the argument of the spin damping term $R_{\rm s}$, however, there is no net contribution of $\Delta m$ to $g_{\rm eff}$ or the spin zero angles (see for example Fig.~\ref{massasymmetry} in methods). Identical values for the effective {\it g}-factor of the $F_-$ and $F_+$ frequencies is therefore a protected property of the inversion symmetry of the bilayer. It can now be understood why the contributions to the beat pattern from $F_-$ and $F_+$ appear to vanish simultaneously at $\theta\approx$~38$^\circ$ in Fig.~\ref{data}a. 

A more elaborate illustration of the predictive power of our physical picture is given by the comparison of the amplitudes of quantum oscillations observed in $c$-axis and in-plane transport (the latter being obtained using the contactless technique~\cite{sebastian1,sebastian2}). As shown in Fig~\ref{FFT}, the amplitudes $A_+$ and $A_-$ are nearly symmetric in quantum oscillations of the in-plane transport whereas they are highly asymmetric in $c$-axis transport measured on the same sample over the same range in magnetic field. A similar difference between $c$-axis and in-plane transport data can also be seen on comparing Refs.~\cite{sebastian2,doiron2}. 

The combination-orbit origin of frequencies $F_-$ and $F_+$ implies that $\frac{3}{4}$ of $A_-$ originates from the antibonding band and the other $\frac{1}{4}$ originates from the bonding band (the fractions being reversed in the case of $A_+$). It has been suggested~\cite{andersen1} that the (inter-unit cell) $c$-axis hybridization can differ significantly for the bonding and antibonding orbitals, with the antibonding orbitals being better hybridized along the $c$-axis. 
A difference between the bonding and antibonding $c$-axis dispersion is also directly evident from the thickness of the Fermi surface contours in Ref.~\cite{elfimov1}. 
This generally implies a higher $c$-axis conductivity for predominantly antibonding orbits (e.g. $F_-$) as compared to predominantly bonding orbits (e.g. $F_+$), in qualitative agreement with observed behavior. The difference in $c$-axis hybridization between bonding and antibonding orbits can be  schematically motivated by the following argument (see Fig.~\ref{FFT}e). The antibonding wave function must vanish in the bilayer-bisecting plane (coincident with the plane containing Y ions in YBa$_2$Cu$_3$O$_{6+x}$). Thus, when compared with bonding orbitals, the charge density in the antibonding orbitals has a larger support {\it outside} the bilayer. A larger support outside the bilayer implies a stronger bilayer-to-bilayer hybridization for the antibonding orbitals. Note that planar transport in Figs.~\ref{FFT}a and c (see also Ref.~\cite{sebastian1} and also the magnetization in Ref.~\cite{audouard1}) is nearly unaffected by these differences in the chemical character of the orbitals.

\section*{Discussion}
We have shown that anomalously small effective {\it g}-factors for the side frequencies originate from combination orbits in which traversal of the nodal region is accompanied by a spin-flip. We have also shown that such a spin-flip is mediated by spin-orbit interactions naturally present in a bilayer system (see Figure~\ref{bilayer}a). The quantum oscillation amplitude of the side frequencies can only be significant if spin-orbit interactions dominate the kinematics in the nodal region despite their weak magnitude in 3$d$ transition metals. A vanishingly small bilayer-splitting in the nodal direction is therefore implied, as originally suggested by Andersen~\cite{andersen1}. We suggest that the residual bilayer splitting observed in Bi$_2$Sr$_2$CaCu$_2$O$_{8+x}$~\cite{kordyuk1} and in YBa$_2$Cu$_3$O$_{6+x}$ at hole dopings below $\approx$~15\%~\cite{fournier1} is directly associated with weak spin-orbit interactions.

The spin-orbit assisted magnetic breakdown combination orbit picture further suggests that similar values of the effective {\it g}-factors for the lower $F_-$ and $F_+$ side frequencies in underdoped YBa$_2$Cu$_3$O$_{6+x}$ are a protected property of the bilayer. Furthermore, differences in the relative amplitudes of the $F_-$ and $F_+$ frequencies in $c$-axis transport compared to in-plane transport (or magnetic torque) can be reconciled with differences in the spatial form of the bonding and antibonding orbitals between bilayers.

The reported value of the characteristic magnetic breakdown nodal energy gap of $\approx$~10~meV~\cite{sebastian1} provides an upper bound for the magnitude of the spin orbit interactions in YBa$_2$Cu$_3$O$_{6+x}$ (due to effects of a small angle in {\it k}-space between the different trajectories as illustrated in Fig.~\ref{trajectories}~\cite{shoenberg1}). The corresponding zero magnetic field momentum space nodal gap $\Delta k_{\rm SO}=\Delta_{\rm SO}/\hbar v_{\rm F}\sim$~0.01~\AA$^{-1}$ (where $v_{\rm F}=\sqrt{2e\hbar F_0}/m^\ast$ is the orbitally-averaged Fermi velocity) is, however, found to be comparable in magnitude to the residual nodal bilayer gap of $\approx$~0.01~\AA$^{-1}$ inferred from photoemission measurements of Bi$_2$Sr$_2$CaCu$_2$O$_{8+\delta}$~\cite{kordyuk1}. It is further consistent with the reported vanishing (below resolution limits) of the residual bilayer splitting in YBa$_2$Cu$_3$O$_{6.5}$~\cite{fournier1} at hole dopings $<$~15~\%. 

We note that  the Stark quantum interference effect~\cite{stark1}, which is a momentum-space analogue of the Aharanov-Bohm effect, can lead to an independent oscillation frequency of $\Delta F$ with a slow temperature decay rate. Such interference effects are expected to be observed in transport measurements, such as the electrical resistivity~\cite{stark1} or themopower~\cite{blanter1}, which is consistent with the recent reports of a slow frequency~\cite{sebastian2,doiron2} similar in value to $\Delta F$ with a very light effective mass. Since quantum interference effects are expected to be absent in static thermodynamic quantities such as the heat capacity and magnetization~\cite{stark1,shoenberg1}, future measurement of these quantities provide a means whereby quantum interference can be distinguished experimentally from alternative explanations, such as an additional small pocket~\cite{doiron2}. 

Further experimental evidence that would substantiate the bilayer spin-orbit picture includes very similar quantum oscillation beat patterns and effective {\it g}-factors observed in other bilayer cuprates such as YBa$_2$Cu$_4$O$_8$~\cite{yelland1,bangura1}. Similar spin-flip phenomena could potentially also occur in bilayer ruthenates~\cite{raghu1,kikugawa1} and heavy fermion superlattices~\cite{shimozawa1}. 
Conversely, the existence of only a single CuO$_2$ plane in HgBa$_2$CuO$_{4+\delta}$~\cite{barisic1} should preclude observation of similar magnetic breakdown combination orbits and anomalously small effective {\it g}-factors in that system.

\section*{Acknowledgements}
This work is supported by 
the US Department of Energy BES ``Science at 100 T" grant no. LANLF100, the National Science Foundation and the State of Florida. We would like to thank Kimberly Putkonen and Ross McDonald for helpful comments.

\section*{Methods}
\footnotesize
\subsection*{Semiclassical bands in the presence of spin-orbit interactions}
\begin{figure}
\centering 
\includegraphics*[width=.5\columnwidth]{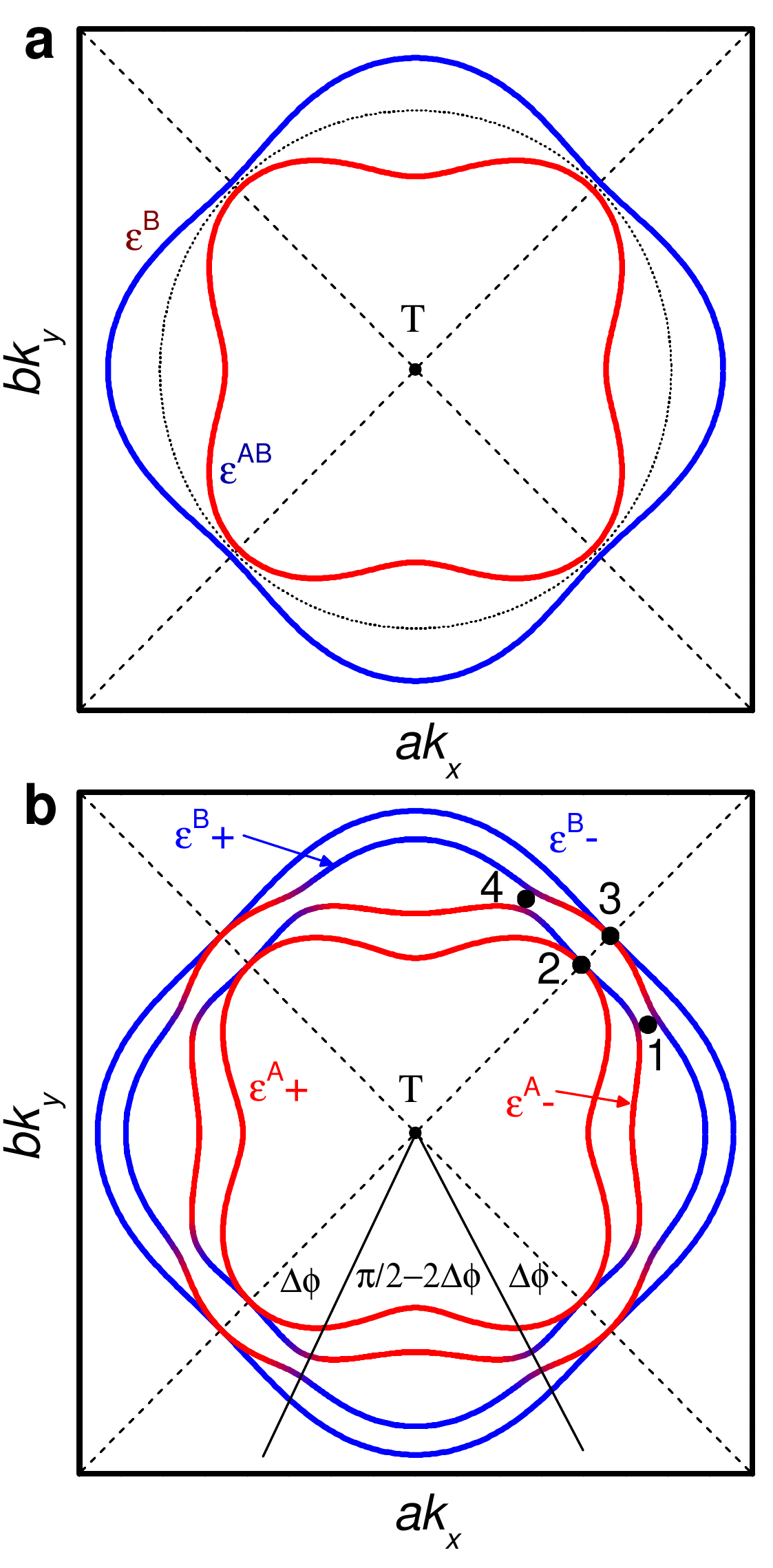}
\caption{{\bf Calculated circular bilayer-split Fermi surface with spin-orbit interactions}. 
({\bf a}), The Fermi surface for $\alpha\neq0$ at $B=0$, with the nodes in $t_{\perp,\phi}=t_{\perp,0}\cos^22\phi$  indicated by dashes lines. The dotted line indicates the Fermi surface for a single layer prior to including spin-orbit interactions and bilayer coupling. ({\bf b}), The same Fermi surface at $B\neq0$ and $\theta=$~0$^\circ$. Magnetic breakdown tunneling junctions are numerically indicated.}
\label{zoom}
\end{figure}

Semiclassically, the primary effect of spin-orbit interactions is to open a gap $\Delta k_{\rm SO}$ between $\Psi_{{\rm A}\uparrow}$ and $\Psi_{{\rm B}\downarrow}$ (see Fig.~\ref{zoom}). As a consequence of sub-leading hopping terms and spin-orbit interactions when $\theta\neq0$, a residual gap $\Delta k_{\rm BA}$ can also open between $\Psi_{{\rm A}\downarrow}$ and $\Psi_{{\rm B}\downarrow}$ and between $\Psi_{{\rm A}\uparrow}$ and $\Psi_{{\rm B}\uparrow}$.

Spin-orbit interactions can be modeled by considering the Fermi surface pocket (which we will assume to be electron-like~\cite{harrison1,sebastian9,doiron2,maharaj1,allais1,zhang1,tabis1,senthil1}) to be located at a high symmetry point intersected by nodes in $t_{\perp,{\bf k}}$~\cite{sebastian1,sebastian2}, such as the T point of a reconstructed body-centered orthorhombic Brillouin zone~\cite{sebastian2} (see Fig.~\ref{zoom}). If we take the center of the pocket as the origin, the bilayer coupling $t_{\perp,\phi}$ can be parameterized as a function of the in-plane polar angle $\phi$. To focus our discussion on the effects of the spin-orbit interactions, we model the single layer electronic dispersion with a simple schematic form $\varepsilon_{\bf k}=\hbar^2k^2/2m^\ast-\mu$ (where $\mu$ is the chemical potential). In the absence of intra-bilayer hopping and spin-orbit interactions, this produces an unperturbed circular Fermi surface of radius $k_0=\sqrt{2m^\ast\mu}/\hbar$ (see dotted line in Fig.~\ref{zoom}a). 

Upon expanding Equation~(\ref{bilayerhamiltonian}) and making the substitution $\frac{1}{2}g\mu_{\rm B}{\bf B}=h(\sin\theta\cos\phi_0,\sin\theta\sin\phi_0,\cos\theta)$ (in which $\phi_0$ is the azimuthal angle of the in-plane magnetic field component), we obtain
\begin{widetext}
\begin{equation}\label{bilayermatrix}
H_{\rm bilayer}=\left( \begin{array}{cccc}\varepsilon_{\bf k}+h\cos\theta&i\alpha k{\rm e}^{i\phi}+h\sin\theta{\rm e}^{i\phi_0}&t_{\perp,\phi}&0\\
-i\alpha k{\rm e}^{-i\phi}+h\sin\theta{\rm e}^{-i\phi_0}&\varepsilon_{\bf k}-h\cos\theta&0&t_{\perp,\phi}\\
t_{\perp,\phi}&0&\varepsilon_{\bf k}+h\cos\theta&-i\alpha k{\rm e}^{i\phi}+h\sin\theta{\rm e}^{i\phi_0}\\
0&t_{\perp,\phi}&i\alpha k{\rm e}^{-i\phi}+h\sin\theta{\rm e}^{-i\phi_0}&\varepsilon_{\bf k}-h\cos\theta\end{array} \right).
\end{equation}
Diagonalization then yields
\begin{eqnarray}\label{bands}
\varepsilon^{\rm B}_\mp=\varepsilon_{\bf k}-\sqrt{(\alpha^2k^2+h^2+t^2_{\perp,0}\cos^42\phi)\pm2h\sqrt{t^2_{\perp,0}\cos^42\phi+\alpha^2k^2\sin^2\theta\sin^2(\phi-\phi_0)}}\nonumber\\
\varepsilon^{\rm A}_\mp=\varepsilon_{\bf k}+\sqrt{(\alpha^2k^2+h^2+t^2_{\perp,0}\cos^42\phi)\mp2h\sqrt{t^2_{\perp,0}\cos^42\phi+\alpha^2k^2\sin^2\theta\sin^2(\phi-\phi_0)}}
\end{eqnarray}
\end{widetext}
for the bonding (B) and antibonding (A) bands, respectively.

For $\theta=0^\circ$, Equation~(\ref{bands}) reduces to $\varepsilon^{\rm B}_\mp=\varepsilon_{\bf k}-\sqrt{(t_{\perp,\phi}\pm h)^2+\alpha^2k^2}$ and $\varepsilon^{\rm A}_\mp=\varepsilon_{\bf k}+\sqrt{(t_{\perp,\phi}\mp h)^2+\alpha^2k^2}$, which yields a simple bilayer split Fermi surface at $h=0$ (i.e. $B=0$, see Fig.~\ref{zoom}a) exhibiting a similar fourfold topology to that shown in Fig.~\ref{bilayer}b and considered Ref.~\cite{sebastian1}. At $h=0$, the residual splitting between the bonding and antibonding Fermi surfaces occurs at $\phi=\pm$~45$^\circ$ and $\pm$~135$^\circ$, and is determined entirely by the strength of the spin-orbit interactions, $\Delta_{\rm SO}=2\alpha k$. 

\subsection*{Transfer amplitudes}
In the semiclassical approximation, the transfer matrices are
\begin{align}\label{TSO}
\hat{T}_{\rm SO}=
\left( \begin{array}{cccc}
 1 & 0 & 0 & 0 \\
 0 & i p_1e^{i\xi_1} & q_1e^{i\xi_1} & 0 \\
 0 & q_1e^{i\xi_1} & i p_1e^{i\xi_1} & 0 \\
 0 & 0 & 0 & 1 \\
\end{array} \right)\,\\
\hat{T}_{\rm BA}=
e^{i\xi_2}\left(
\begin{array}{cccc}
 q_2 & 0 & i p_2 & 0 \\
 0 & q_2& 0 & i p_2 \\
 i p_2 & 0 & q_2& 0 \\
 0 & i p_2 & 0 & q_2 \\
\end{array}\right),
\end{align}
where $\xi_1$ and $\xi_2$ represent phases that could potentially contribute to the quantum oscillation phase~\cite{shoenberg1}. The amplitude for a particular complete orbit is given by a product of the transfer matrices for each juntion, interspersed by the amplitudes describing free propagation of an electron in the crystal under the action of the Lorentz force between the nodes along one of the four trajectories. The free propagation can be described by a diagonal matrix $\hat{F}_\Phi=\text{diag}( e^{i\psi_{{\rm B}\uparrow}}, e^{i\psi_{{\rm B}\downarrow}}, e^{i\psi_{{\rm A}\uparrow}}, e^{i\psi_{{\rm A}\downarrow}} )$ where $\psi_{{\rm B}\uparrow}=\Phi\big(\frac{F_0+\frac{1}{2}\Delta F}{B\cos\theta}+\frac{(m+\frac{1}{2}\Delta m)g}{4m_{\rm e}\cos\theta}\big)$, $\psi_{{\rm B}\downarrow}=\Phi\big(\frac{F_0+\frac{1}{2}\Delta F}{B\cos\theta}-\frac{(m+\frac{1}{2}\Delta m)g}{4m_{\rm e}\cos\theta}\big)$,  $\psi_{{\rm A}\uparrow}=\Phi\big(\frac{F_0-\frac{1}{2}\Delta F}{B\cos\theta}+\frac{(m-\frac{1}{2}\Delta m)g}{4m_{\rm e}\cos\theta}\big)$, $\psi_{{\rm A}\downarrow}=\Phi\big(\frac{F_0-\frac{1}{2}\Delta F}{B\cos\theta}-\frac{(m-\frac{1}{2}\Delta m)g}{4m_{\rm e}\cos\theta}\big)$ are the phase factors and $\Phi$ is the polar angle in momentum-space between junctions (relative to the center of the orbit). Between junctions 1 and either 2 or 3 (or between junctions 2 or 3 and 4), $\Phi=\Delta\phi$, while between junctions 1 and 4 at adjacent nodes, $\Phi=\frac{\pi}{2}-2\Delta\phi$. Meanwhile, $\Delta m$ allows for a possible difference in cyclotron mass between bonding and antibonding portions of the orbit (see below). On expanding the trace
${\rm Tr}[(\hat{T}^{\dagger}_{\rm SO}\hat{F}_{\Delta\phi}\hat{T}_{\text{BA}}\hat{F}_{\Delta\phi}\hat{T}_{\rm SO}\hat{F}_{\frac{\pi}{2}-2\Delta\phi})^4]$ (where the 4$^{\rm th}$ power accounts for traversal through 4 consecutive nodal regions), we obtain a series of terms $\sum_iA_ie^{i\frac{2\pi F_i}{B\cos\theta}+\frac{\pi m_{{\rm c}}g_{{\rm eff},i}}{2}}$. The magnetic breakdown amplitude factor $R_{\rm MB}$ for each type of orbit is obtained by collecting terms of the same frequency $F_i$ and effective {\it g}-factor $g_{{\rm eff},i}$ and summing their amplitudes $A_i$.
In the limits $\Delta m\rightarrow0$ and $\Delta\phi\rightarrow0$, for $F_0$ we obtain:
\begin{align}\label{F0}
g_{\rm eff}=2,\qquad& \frac{1}{2} e^{4 i (\xi_1 +\xi_2 )} p_2^2 p_1^2 \left(2 q_2^2 \left(q_1^2-p_1^2\right)+p_2^2 p_1^2\right)\notag\\
g_{\rm eff}=1, \qquad&  4 e^{2i(3 \xi_1+2 \xi_2) } p_2^2 q_2^2 p_1^2 q_1^2 \left(1-3 p_1^2\right)\notag\\
g_{\rm eff}=0, \qquad& 2 e^{4 i (\xi_1 +\xi_2 )} p_1^2 q_1^2 \left(-p_2^4-2 e^{4 i \xi_1 } q_2^4 \left(1-6 p_1^2 \right)\right)\notag\\
g_{\rm eff}=-1, \qquad& 4 e^{2i(3\xi_1+2\xi_2)} p_2^2 q_2^2 p_1^2 q_1^2 \left(1-3 p_1^2\right)\notag\\
g_{\rm eff}=-2, \qquad& \frac{1}{2} e^{4 i (\xi_1 +\xi_2 )} p_2^2 p_1^2 \left(2 q_2^2 \left(q_1^2-p_1^2\right)+p_2^2 p_1^2\right),
\end{align}
for $F_-$ we obtain:
\begin{align}\label{F1}
g_{\rm eff}=2, \qquad& e^{2 i (\xi_1 +2\xi_2 )} p_2^2 p_1^2 q_2^2\notag\\
g_{\rm eff}=1, \qquad&  -e^{4 i (\xi_1 +\xi_2 )} p_2^2 p_1^2 q_1^2 \left(p_2^2-4 q_2^2\right)\notag\\
g_{\rm eff}=0, \qquad& -4 e^{2i(3\xi_1+2\xi_2)} p_2^2 q_2^2 q_1^2 \left(p_1^4-2 p_1^2 q_1^2\right)\notag\\
g_{\rm eff}=-1,  \qquad& e^{4 i (\xi_1 +\xi_2 )} p_1^2 q_1^2 \left(-p_2^4-4 e^{4 i \xi_1 } q_2^4 \left(p_1^2-q_1^2\right){}^2\right) \notag\\
g_{\rm eff}=-2, \qquad& e^{2i(3\xi_1+2\xi_2)} p_2^2 p_1^2 q_2^2 \left(p_1^2-q_1^2\right){}^2,
\end{align}
while for $F_{--}$ we obtain:
\begin{align}\label{F2}
g_{\rm eff}=2, \qquad& \frac{1}{4} e^{4 i \xi_2} q_2^4\notag\\
g_{\rm eff}=1, \qquad& e^{2 i (\xi_1 +2\xi_2 )} p_2^2 q_2^2 q_1^2\notag\\
g_{\rm eff}=0, \qquad& \frac{1}{2} e^{4 i (\xi_1 +\xi_2 )} p_2^2 q_1^2 \left(2 p_1^2 q_2^2+q_1^2 \left(p_2^2-2 q_2^2\right)\right)\notag\\
g_{\rm eff}=-1, \qquad& e^{2i(3\xi_1+2\xi_2)} p_2^2 q_2^2 q_1^2 \left(p_1^2-q_1^2\right){}^2\notag\\
g_{\rm eff}=-2, \qquad& \frac{1}{4} e^{4 i (2\xi_1+\xi_2 )} q_2^4 \left(p_1^2-q_1^2\right){}^4
\end{align}
for the effective {\it g}-factors and amplitudes, respectively. Here, positive and negative signs for $g_{\rm eff}$ refer to orbits for which the net spin projection is respectively up and down. Equal amplitudes for for each spin component are additive. In the case of unequal amplitudes, the quantum oscillation phase contributions from the spin-up and spin-down components no longer destructively interfere at each spin zero. For $F_+$ and $F_{++}$ we obtain the same amplitudes as for $F_-$ and $_{--}$, but with the corresponding signs of $g_{\rm eff}$ reversed. An examination of Equations (\ref{F0}) through (\ref{F2}) reveals that there is no relative phase between any of the dominant quantum oscillation frequencies, allowing us to neglect the phase by setting $\xi_1=\xi_2=0$ in the main text, as is commonly done with conventional magnetic breakdown networks. Non zero values of $\xi_1$ and $\xi_2$ can only contribute to the overall phase of the quantum oscillations.

\subsection*{Chemical difference between B and A}
\begin{figure}[ht!!!!!]
\centering
\includegraphics[width=1\columnwidth]{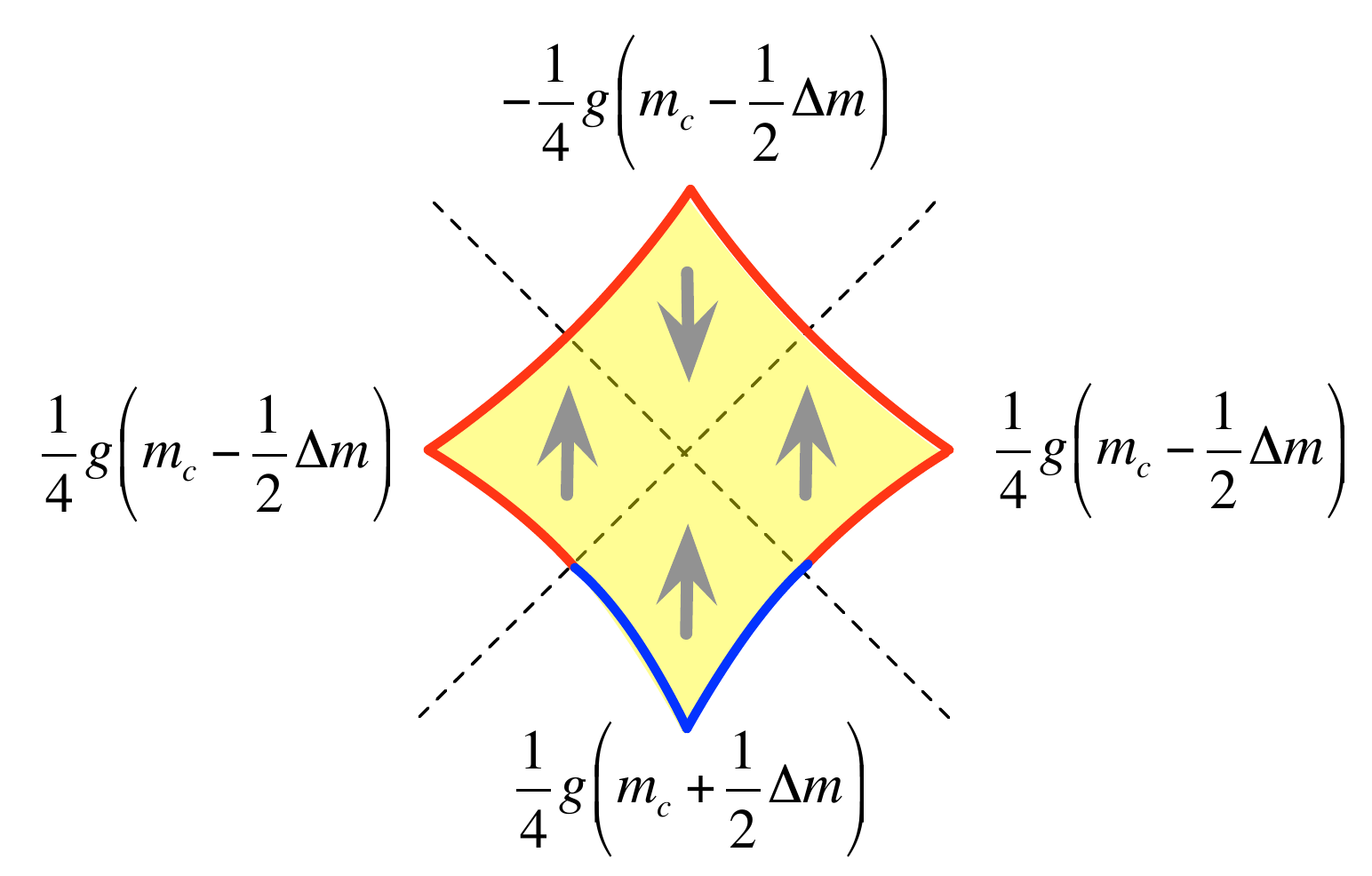}
\caption{{\bf Effect of a difference in effective mass $\Delta m$ between B and A bands}. 
A schematic showing the contributions to the product $mg_{\rm ff}$ from bonding and antiboding bands for each of the orbit quandrants, with the difference in effective masses parameterized by $\Delta m$.
}
\label{massasymmetry}
\end{figure}

On including the effect of a difference in effective mass $\Delta m$, the product $mg_{\rm eff}$ that enters into the argument of $R_{\rm s}$ is obtained by summing contributions from each of the orbit quadrants. Figure~\ref{massasymmetry} shows such a summation for the case of the orbit yeilding the $F_-$ side frequency, where it is found that the product $mg_{\rm eff}=2\times\frac{1}{4}(m-\frac{1}{2}\Delta m)g+\frac{1}{4}(m+\frac{1}{2}\Delta m)g-\frac{1}{4}(m-\frac{1}{2}\Delta m)g=\frac{1}{2}mg$ is independent of $\Delta m$. A similar result is found for all orbits, and on considering a difference in $g$ between the bonding and antibonding bands instead of the effective mass.


\subsection*{Semiclassical orbits at large $\theta$}
When the magnetic field is rotated into the planes, spin-orbit coupling is found to contribute to the lifting of the nodal degeneracy between $\varepsilon^{\rm B}_+$ and $\varepsilon^{\rm A}_+$ and between $\varepsilon^{\rm B}_-$ and $\varepsilon^{\rm A}_-$ in Equation~(\ref{bands}). This does not affect the point of intersection between between $\Psi_{A\uparrow}$ and $\Psi_{B\downarrow}$ in Fig.~\ref{trajectories} and therefore does not affect the effective {\it g}-factors. It does, however, cause $q_2$ to increase at high angles. Numerical simulations indicate that a non-zero $q_2$ attributed only to spin-orbit interactions introduces new orbits of the same frequencies (i.e. $F_{--}$~\dots~$F_{++}$), many of which have $g_{\rm eff}\approx2$. Their overall amplitude is found to remain below  $\sim$~20~\%~of the dominant amplitude $A_0$ at $\theta\approx$~70$^\circ$.


Author contribution statement: N.H. and A.S. wrote and reviewed the manuscript and prepared the figures. B.J.R. contributed data for the figures and reviewed the manuscript.\\

Additional information: The author(s) declare no competing financial interests.

\end{document}